# Redshifts vs Paradigm Shifts: Against Renaming Hubble's Law

Cormac O'Raifeartaigh and Michael O'Keeffe

*School of Science and Computing, Waterford Institute of Technology, Cork Road, Waterford, Ireland.*

Author for correspondence: coraifeartaigh@wit.ie

**Abstract**

We consider the proposal by many scholars and by the International Astronomical Union to rename Hubble's law as the Hubble-Lemaître law. We find the renaming questionable on historic, scientific and philosophical grounds. From a historical perspective, we argue that the renaming presents an anachronistic interpretation of a law originally understood as an empirical relation between two observables. From a scientific perspective, we argue that the renaming conflates the redshift/distance relation of the spiral nebulae with a universal law of spatial expansion derived from the general theory of relativity. We note that the first of these phenomena is merely one manifestation of the second, an important distinction that may be relevant to contemporary puzzles concerning the current rate of cosmic expansion. From a philosophical perspective, we note that many of the named laws of science are empirical relations between observables, limited in range, rather than laws of universal application derived from theory.

## 1. Introduction

In recent years, many scholars[1] have suggested that the moniker Hubble's law – often loosely understood as a law of cosmic expansion – overlooks the seminal contribution of the great theoretician Georges Lemaître, the first to describe the redshifts of the spiral nebulae in the context of a cosmic expansion consistent with the general theory of relativity. Indeed, a number of authors[2] have cited Hubble's law as an example of Stigler's law of eponymy, the assertion that "no scientific discovery is named after its original discoverer".[3] Such scholarship recently culminated in a formal proposal by the International Astronomical Union (IAU) to rename Hubble' law as the "Hubble-Lemaître law". The IAU proposal was advanced at the 30th meeting of the Union in August 2018 and had four stated aims:[4]

> (i) to pay tribute to both Georges Lemaître and Edwin Hubble for their fundamental contributions to the development of modern cosmology
> (ii) to honour the intellectual integrity of Georges Lemaître that made him value more the progress of science rather than his own visibility
> (iii) to highlight the role of the IAU General Assemblies in fostering exchanges of views and international discussions
> (iv) to inform the future scientific discourses with historical facts

Members of the IAU worldwide voted electronically on the resolution on October 26th 2018 and it was passed with a majority of 78%. This decision was reported uncritically in many media outlets[5] and a process of renaming appears is well under way; for example, the new nomenclature appears in the Wikipedia entry on Hubble's law.[6] Curiously, there has been almost no discussion of this decision in the historical literature to date.[a]

In this article, we will argue that the renaming of Hubble's law is questionable on historic, scientific and philosophical grounds. From a historical perspective, we will argue that the renaming is quite revisionist as it overlooks the fact that Hubble's law was understood for many years as an empirical relation obeyed by certain astronomical bodies. From a scientific perspective, we will argue that the renaming conflates two distinct advances, Edwin Hubble's discovery of a linear relation between redshift and distance for the nebulae and Georges Lemaître's derivation of a universal law of spatial expansion from the general theory of relativity. From a philosophical perspective, we will argue that the renaming overlooks the fact that many of the named laws of science are empirical in nature.

---
[a] We are aware of one preprint on the topic, discussed in section 4.

**2. Historical background : a slow paradigm shift**

In 1929, the well-known American astronomer Edwin Hubble published the first results of a detailed investigation of the relation between the redshifts of the nebulae and their distance,[7] using redshift data from V.M. Slipher and newly-derived estimates of nebular distances based on Hubble's recent observations of individual stars within the nebulae. The results are reproduced in figure 1. As stated by Hubble: *"the results establish an approximately linear relation between the velocities and distances among nebulae for which velocities have been previously published"*. We note that the distances of the closest seven nebulae were estimated by observing Cepheid stars within the nebulae and employing Henrietta Leavitt's period-luminosity relation to estimate their distance; the next thirteen distances were estimated by observing the most luminous stars in nebulae and assuming an upper limit of absolute magnitude $M = -6.3$; the remaining four objects had distances assigned on the basis of the mean luminosities of the nebulae in a cluster. Finally, the single cross represents a mean velocity/distance ratio for 22 nebulae whose distances could not be estimated individually. Many commentators[8] have noted that the data shown on Hubble's graph only marginally supported the conclusion of a linear relation between redshift and distance for the nebulae. However, the graph marked an important turning point as the data were accepted as the first evidence of a relation that had long been suspected by astronomers.[9] This conclusion was strengthened with the publication of a paper soon afterwards that extended the results to much larger distances and redshifts.[10]

On January 10th 1930, a landmark meeting took place at the Royal Astronomical Society in Burlington House in London. Prominent astronomers such as Arthur Stanley Eddington and Willem de Sitter noted that the recent observations of an approximately linear relation between the redshifts of the spiral nebulae and their radial distances could not be readily explained in the context of the standard mathematical models of the cosmos, i.e., in the context of the static cosmology of Albert Einstein or the empty cosmology of de Sitter. During the course of the meeting, it was suggested that non-static cosmologies should be considered.[11] This discussion was reported in the February issue of *The Observatory*[12] and read by Georges Lemaître, who immediately wrote a letter to Eddington, reminding him of his own 1927 article[13] on this very topic. As Lemaître wrote: "*Dear Professor Eddington, I just read the February $N^0$ of the Observatory and your suggestion of investigating of non-statical intermediary solutions between those of Einstein and de Sitter. I made these investigations two years ago. I consider a universe of curvature constant in space but increasing in time. And I emphasize the existence of a solution in which the motion of the nebulae is always a receding one from time minus infinity to plus infinity*".[14] As is well-known, Eddington responded with enthusiasm, publicly acknowledging the importance of Lemaître's 1927 article and bringing it to the attention of his colleagues. Eddington also ensured the article reached a wide audience by arranging for it to be republished in English in the *Monthly Notices of the Royal Astronomical Society*.[15]

There is little question that the kernel of Lemaître's 1927 paper was his brilliant hypothesis of a connection between the redshifts of the spiral nebulae and an expansion of space derived from the general theory of relativity. Indeed, it is generally accepted that it is this aspect of his paper that distinguishes his contribution from that of Alexander Friedman.[16]

However, Lemaître's paper was not primarily concerned with astronomical observations; instead he predicted a linear relation between redshift and distance for distant astronomical bodies from his hypothesis of cosmic expansion, and employed a mean value of the redshift/distance ratio for the nebulae (as far as it was known) to extract a rough estimate of the rate of expansion. It's worth noting that the astronomical data cited in this paper were much more tentative than Hubble's graph of 1929, a point that is often overlooked.[17]

With the publication of Hubble's detailed observations of the nebulae in 1929, and the republication of Lemaître's theory in English in 1931, it seemed to many theorists that the redshifts of the nebulae could be explained in terms of a cosmic expansion predicted by the general theory of relativity. By the early 1930s, a number of articles by leading theorists on expanding cosmologies with varying cosmic parameters had been published.[18] Even Einstein overcame his earlier distrust of time-varying models of the cosmos and proposed two dynamic cosmic models during this period, the Friedman-Einstein model of 1931 and the Einstein-de Sitter model of 1932.[19]

However, such changes in scientific worldview are rarely instant or unanimous, as pointed out by philosophers of science such as Thomas Kuhn[20] and Imre Lakatos.[21] Certainly, Hubble's redshift/distance data were soon accepted, despite some initial objections from astronomers such as Harlow Shapley.[22] One reason was undoubtedly the use of Cepheid variables to measure the distances of some nebulae. Hubble's status as a leading astronomer working at the world's largest telescope may also have played a role.[23] A third factor was the publication of similar data for nebulae at much greater distance in the years that followed.[24] By the mid-1930s, few doubts remained concerning the validity of the redshift/distance relation.

By contrast, the interpretation of Hubble's data in terms of cosmic expansion was far from settled in this period. One obvious problem was that most expanding models seemed to predict an age for the universe that was problematic. Many theoreticians noted that for the simplest models, Hubble's graph implied a universe that had been expanding for about two billion years. This was a curious figure if it represented the age of the cosmos, since experiments from radioactivity suggested that the earth was at least four billion years old![25] Thus, several alternative explanations for the recession of the nebulae were offered in these years. The best known of these was a hypothesis from the Swiss physicist Fritz Zwicky that light from distant stars might be redshifted due to a loss of energy as it travelled over vast distances in interstellar space.[26] Indeed, quite a number of physicists made similar suggestions, a class of theories that became known as 'tired-light' theories.[27] Another hypothesis was that the redshifts of the nebulae represented a Doppler effect due to the movement of galaxies into neighbouring space, a suggestion that was advanced in the context of the so-called kinematic cosmology of Edward Milne.[28] Thus, in the 1930s many astronomers and theorists kept an open mind regarding the meaning of the redshifts.[29] As Hubble and Richard Tolman remarked in a joint publication[30] on observational and theoretical investigations of the nebulae:

> Until further evidence is available, both the present writers wish to
> express an open mind with respect to the ultimately most satisfactory
> explanation of the nebular red-shift and, in the presentation of
> purely observational findings, to continue to use the phrase "apparent"
> velocity of recession. They both incline to the opinion, however,
> that if the red-shift is not due to recessional motion, its explanation
> will probably involve some quite new physical principles.

As the years progressed, non-relativistic explanations for the redshift/distance relation of the nebulae were eventually ruled out.[31] However, some astronomers, including Hubble himself, remained agnostic on the subject throughout their careers. Hubble's attitude is perhaps most clearly seen in his last public address, the 1953 Darwin Lecture of the Royal Astronomical Society:[32]

> I propose to discuss the law of red-shifts—the correlation
> between distances of nebulae and displacements in their
> spectra. It is one of the two known characteristics of the
> sample of the universe that can be explored and it seems to
> concern the behaviour of the universe as a whole. For this
> reason it is important that the law be formulated as an
> empirical relation between observed data out to the limits

> of the greatest telescope. Then, as precision increases, the array of possible interpretations permitted by uncertainties in the observations will be correspondingly reduced. Ultimately, when a definite formulation has been achieved, free from systematic errors and with reasonably small probable errors, the number of competing interpretations will be reduced to a minimum.

Hubble's failure to embrace the thesis of cosmic expansion has become the subject of much comment in recent years, and it is certainly somewhat ironic in the context of modern nomenclature such as the *Hubble flow* and the *Hubble constant.* However, it should be borne in mind that his careful demarcation between observation and explanation was not uncommon amongst astronomers at the time, particularly in cases where the explanation involved abstruse theories such as the general theory of relativity.[33] Indeed, it could be stated that astronomers of this period were engaged in 'cosmology by accident rather than design', as noted by the historian Robert Smith.[34] We also note that it was not until the mid-1950s that the time-scale difficulty associated with expanding cosmologies was resolved, due to important revisions in the cosmological distance scale.[35]

## 3. The naming of Hubble's law

In time, the graph of figure 1 became known as 'Hubble's law'. It is not entirely clear when this nomenclature became the norm. Certainly, there are copious references to 'Hubble's observations' and 'Hubble's relation' in the cosmological papers of the early 1930s cited in section 2. By 1933, at least two specific references to 'Hubble's law' had appeared in the literature, in papers by Edward Arthur Milne[36] and by Arthur Geoffrey Walker.[37] Both theoreticians were well-known and their nomenclature may have been influential. However, the use of the moniker 'Hubble's law' only seems to have become widespread in the 1950s, possibly through its use in popular books such as *'The Creation of the Universe'* by George Gamow,[38] *'The Unity of the Universe'* by Denis Sciama [39] and *'The Expansion of the Universe'* by Pierre Couderc.[40] Indeed, it is interesting to note that Lemaître himself employed the nomenclature "*la loi de Hubble*" in his review of the French edition of the latter book, as discussed below.

**4. On the renaming of Hubble's law**

*Historical considerations*

We recall first that Hubble's law was understood for many years as an empirical relation between the redshifts and the distances of the spiral nebulae (see section 3 above). Thus, from a historical perspective, it is quite anachronistic to include Lemaître in this nomenclature. Lemaître did not provide any measurements of redshift or distance of the nebulae, nor did he establish the linearity of the redshift/distance relation. Instead, he *predicted* a linear relation between redshift and distance as a manifestation of a general expansion of space derived from relativistic cosmology; assuming such a relation existed, he estimated an average co-efficient of cosmic expansion from preliminary values for redshift and distance taken from the observational data of Slipher and Hubble. That this calculation was something of a provisional 'guesstimate' can be seen from the fact that it appears only as a single line in the republished version of the paper.[41] Indeed, Lemaître's own attitude towards the astronomical data he used in 1927 is made clear in the covering letter that accompanied his 1931 translation: *"I do not think it is advisable to reprint the provisional discussion of radial velocities which is clearly of no actual interest"*.[42] The same attitude can be seen in a comment made by Lemaître many years later in a review[43] of Pierre Couderc's book: *"Naturellement, avant la decouverte et l'etude des amas de nebuleuses, il ne pouvait être question d'établir la loi de Hubble"* or *"Naturally, before the discovery and study of the clusters of nebulae, there could be no question of establishing Hubble's law"*. Similarly, in a review article written in 1952, Lemaître wrote: "*Hubble and Humason established from observation the linear relation between velocity and* *distance which was expected for theoretical reasons and which is known as the Hubble velocity-distance relation"*.[44]

Thus, the renaming of Hubble's law as the Hubble-Lemaître law can be seen as an example of *presentism*, that is, the anachronistic introduction of latter-day ideas and perspectives into depictions of the past. Indeed, historians have long noted that practitioners of science have a tendency to write presentist or Whiggish histories of their subject.[45] While some might argue that Lemaître interpreted the redshifts of the nebulae in terms of a cosmic expansion as early as 1927, this interpretation was not accepted by the physics community for some time. Meanwhile, the moniker Hubble's relation or Hubble's law was used to denote Hubble's observations of the nebulae for many years.

*Scientific considerations*

Historical considerations aside, the renaming of Hubble's law is also questionable from a scientific perspective. This is because it conflates two distinct phenomena, an empirical relation between redshift and distance observed for certain astronomical bodies and a universal law of spatial expansion derived from the general theory of relativity. (This point has also been made in a preprint by the historian Helge Kragh).[46] Indeed, in our view the proposal confuses *astronomy*, the study of the stars, with *cosmology*, the study of the universe as a whole. As noted by the astronomer Edward Harrison:[47]

> The redshift-distance law $zc = HL$…is the observers' linear law first established by Slipher's redshift measurements and Hubble's distance determinations. Its proper name is the *Hubble law*. From the time of its discovery most cosmologists have realised that in its linear form it is only approximately true. On the other hand, the velocity-distance law $V=HL$…is the theorists' linear law that follows automatically from the assumption that expanding space is uniform (isotropic and homogeneous). This law, often improperly referred to as the Hubble law, is of central importance in modern cosmology and is rigorously true in all uniform universes.

One obvious distinction between the two phenomena is that a simple relation between redshift and distance is not observed in the case of nebulae (or other astronomical bodies) at relatively close distance, as the effects of cosmic expansion are overwhelmed by local gravitational effects. Indeed, this phenomenon hampered the early investigations of the redshift/distance relation of astronomical bodies.[48] More generally, a clear distinction should be drawn in principle between a cosmological redshift, caused by the stretching of space, and a Doppler shift of wavelength due to motion; indeed, some celestial bodies can exhibit both effects simultaneously.[49]

*Philosophical considerations*

Finally, we also find the renaming of Hubble's law questionable from the point of view of the philosophy of science. As many philosophers of science have pointed out, one can distinguish between laws of science that are empirical relations between observables, limited in range, and laws of universal application derived from theory. As the philosopher Peter Caws put it:[50]

> The distinction between hypotheses and empirical generalizations suggests a distinction between two different kinds of scientific law, one corresponding to empirical generalizations which are accepted as true and the other to hypotheses which are accepted as true. In the latter category would fall, for instance, the law of conservation of energy; energy is not observed, but rather the penetration of bullets or the compression of strings, so that any statement about it must be hypothetical.

Thus *Ohm's law* is not a general law of nature, but an empirical relation between current and voltage observed to hold in some materials. Similarly, *Boyle's law* is an empirical relation between pressure and volume obeyed by most gases at constant temperature. Many other laws of science are really empirical relations, from Coulomb's law of electrostatics to Snell's laws of refraction.[51]

In this context, Hubble's redshift/distance relation is seen as an empirical relation, rather than a universal law, and should not be confused with a general law of spatial expansion derived from relativistic cosmology. The two are not equivalent, not least because the former is of limited validity, as pointed out above. It's also worth noting that the redshifts of the nebulae are merely one manifestation of cosmic expansion; other manifestations exist, notably the frequency range of the cosmic microwave background (CMB). Indeed, studies of the CMB provide an alternative measurement of cosmic expansion and it is one of the great puzzles of contemporary cosmology that there is some discrepancy between estimates of the rate of cosmic expansion derived from redshift/distance observations of type Ia supernovae and estimates of the rate of expansion derived from measurements of the cosmic background radiation.[52] The source of the discrepancy remains unclear at the time of writing, but the puzzle serves as a useful reminder that an empirical relation between redshift and distance for certain celestial bodies is merely one manifestation of cosmic expansion.

**Conclusions**

In our view, the renaming of Hubble's law does not represent good historical practice as it presents an anachronistic interpretation of the original meaning of the law. It also doesn't represent good scientific practice as it conflates two distinct phenomena, a linear relation between redshift and distance observed for certain astronomical bodies and a universal law of spatial expansion derived from the general theory of relativity. Indeed, the former constitutes only one manifestation of the latter. We note in addition that many of the named laws of science are empirical in nature.

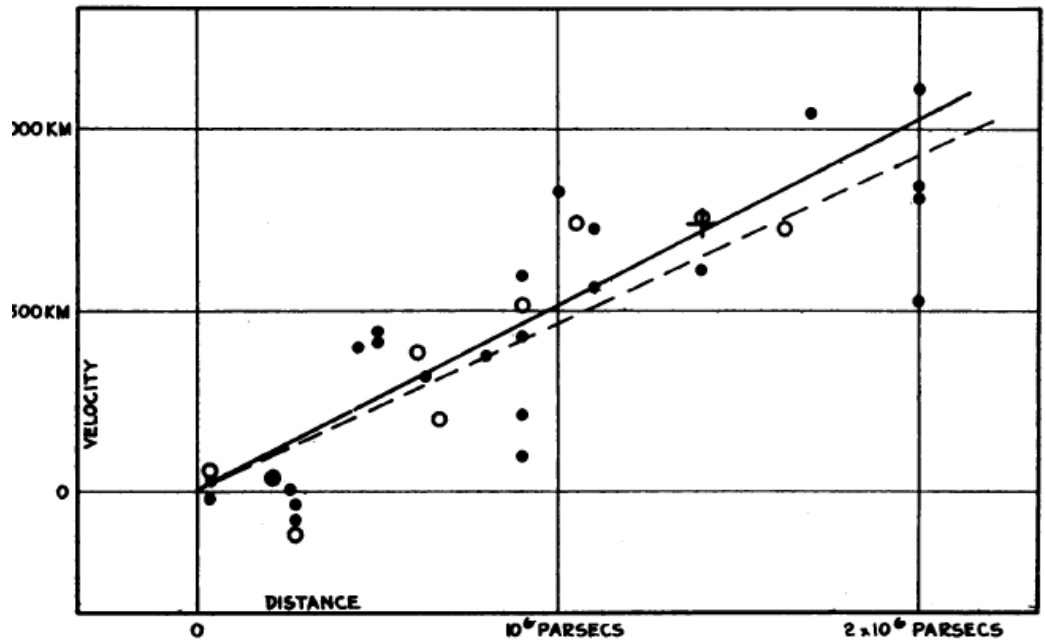

**Figure 2.** Graph of apparent velocity vs distance for the spiral nebulae (reproduced from Hubble ref. 7). Filled circles represent data where solar motion was corrected for individual nebulae; open circles represent data where solar motion was corrected for nebulae in groups.